\documentclass[10pt,aps,prd,twocolumn,showpacs,showkeys,floatfix]{revtex4}

\usepackage{amsmath,amssymb}
 \usepackage{epsfig}
\usepackage{graphicx}% Include figure files
\usepackage{bm}
\usepackage{amsmath}
\newcommand{\beq}{\begin{equation}}
\newcommand{\eeq}{\end{equation}}
\newcommand{\beqa}{\begin{eqnarray}}
\newcommand{\eeqa}{\end{eqnarray}}
\newcommand{\beqar}{\begin{eqnarray*}}
\newcommand{\eeqar}{\end{eqnarray*}}

\begin{document}

\title{Regular black holes: Guilfoyle's electrically charged solutions
with a perfect fluid phantom core}

%\title{Regular black holes: Electrically charged
%solutions, Reissner-Nordstr\"om outside a Guilfoyle-phantom core}

\author{Jos\'e P. S. Lemos}
\affiliation {Centro Multidisciplinar de Astrof\'{\i}sica (CENTRA),
Departamento de F\'{\i}sica,  Instituto Superior T\'ecnico (IST),
Universidade de Lisboa (UL),
Avenida Rovisco Pais 1, 1049-001 Lisboa, Portugal
\\Email: joselemos@ist.utl.pt}

\author {Vilson T. Zanchin}
\affiliation {Centro de Ci\^encias Naturais e Humanas, Universidade
Federal
do ABC,  Avenida dos Estados 5001, Santo Andr\'e,  09210-580 S\~ao Paulo,
Brazil \\ Email: zanchin@ufabc.edu.br}

%\date{today}
%\date{today}

\begin{abstract}

Regular black hole solutions are found among the Guilfoyle exact
solutions. These are spherically symmetric solutions
of general relativity coupled to Maxwell's electromagnetism
and charged matter where the metric potentials
and electromagnetic fields are related in some particularly simple
form.  We show that, for certain ranges of the parameters,
there are objects which correspond to regular charged black holes,
whose interior region is filled by an electrically 
charged phantomlike fluid, or,
in the limiting case, a de Sitter false vacuum 
fluid, and whose exterior region is
Reissner-Nordstr\"om. The boundary between both regions is a smooth
boundary surface, except in the limiting case where the boundary is
made of a massless electrically charged spherically symmetric
coat. The main physical and geometrical properties of such charged
regular solutions are analyzed.

\end{abstract}

%\begin{keyword}
\keywords{Regular black holes; null and timelike boundaries;
Einstein-Maxwell equations}
%\end{keyword}

\pacs{04.70.Bw, 04.20.Jb, 04.40.Nr, 04.40.Dg}

%\keywords{Quasiblack holes, Black holes,
%Classical theories of gravity}

\maketitle

%\end{frontmatter}

%\date{today}

\section {Introduction}
\label{sec-introd}

Solutions of Einstein-Maxwell with matter equations
are of importance as they not only represent 
a class of solutions in general relativity but
also give hints to the behavior 
of solutions with other
type of fields.
Electric charged compact objects, starlike,
solutions have been constructed by Guilfoyle in 
\cite{guilfoyle}. The 
interpretation of the fluid content 
in these solutions has appeared in \cite{Lemos:2009mr},
and part of the spectrum of the solutions
received a global interpretation in 
\cite{lemoszanchin2010,lemoszanchin2015}.

On the other hand, a class of compact objects that has been of
interest are regular black holes, i.e., black holes that, contrary to
the usual Schwarzschild and Reissner-Nordstr\"om black holes, do not
possess curvature singularities at the center.  In these regular black
holes a well-behaved interior replaces the singularity.

The first work on regular black 
holes is due to Bardeen \cite{bardeen1968}
whose regular black hole was put on a firm footing in \cite{ab00}.
Dymnikova \cite{dy92} constructed a regular black hole solution whose
interior matter is an anisotropic fluid obeying a de Sitter equation
of state, $p=-\rho$, where $\rho$ and $p$ are the energy density and
pressure of the fluid, respectively, that asymptotically approached the
Schwarzschild spacetime.  Other regular black hole solutions have been
found in \cite{Lemos:2011vz,yosh} where the interior region is a de
Sitter core that can be matched to a Reissner-Nordstr\"om exterior metric
thanks to an energyless charged layer at the boundary surface
(see also \cite{barisrl91}).  Regular
black holes with nonlinear electrodynamics sources appear in, e.g.,
\cite{mat08,bala,ma}. Stability of these black holes appeared in
\cite{flachilemos,tosh}.  Regular black holes in different theories of
gravity also appeared in \cite{good,rjmz}, and in nonminimal theories
in \cite{balaklemoszaya}.

An important class of regular black holes is the class of phantom
matter black holes.  
Phantom matter is characterized by a perfect fluid for
which $\rho +p<0$.  Fluids satisfying such a condition may represent
the dark energy sector of the Universe.  The property of phantom
matter to act as a negative effective mass turns it into a natural
candidate to build models of compact objects and, in particular, to
build black hole solutions free from singularities.  This motivated
several works reporting on black holes with phantom matter, some of
them regular, see
\cite{br06,br071,c1,a1,br12,br12b,brbook,eiroa,akbar,a2}.

Now, electric charged matter provides repulsion, but in general 
this electric repulsion is not enough to furnish regularity at the
center.  However, electrically charged matter made of a phantom perfect
fluid might indeed be suitable to produce enough antigravity in order to
form a regular core. Thus it is worth exploring the possibility that the
Guilfoyle solutions \cite{guilfoyle} contain a subset of regular electrical
black holes with a phantom perfect fluid core.

In the present work we analyze in detail the Guifoyle model for
electrically charged matter \cite{guilfoyle} and show that within a
certain range of the parameters of the model there are indeed regular
black holes whose central core is made by a charged perfect fluid
obeying an equation of state representing phantom matter for which $p
+\rho<0$.

The layout of the paper is as follows.  In
Sec.~\ref{sec-sphericalGuilfoyle0} we present the basic equations
describing a spherically symmetric electrically charged perfect fluid
matter distribution satisfying some simplifying assumptions, that
define the systems.  We give explicitly the interior and exterior
solutions with the appropriate boundary conditions and display the
class of the Guilfoyle solutions that is of interest here. We also
show the conditions on the solutions necessary to make them black
holes. Section~\ref{sec-rbhgen} is dedicated to analyzing the phantom
regular charged black hole solutions we have found,  presenting their
main properties, and comparing the present solutions to other regular
black hole solutions found in the literature. In
Sec.~\ref{sec-regdeSitterC} a particular limit of the regular black
holes analyzed, where the core is made of de Sitter material and there
is a massless charged layer at the Cauchy horizon, is studied. Then in
Sec.~\ref{sec-analysis} the spectrum of the free parameters of the
solutions that yields regular black holes is analyzed in detail. In the
first part, we study the behavior of the mass and other free
parameters in terms of the radius and energy density of the solutions.
The second part contains the display of the regions of the parameter
space where regular black holes can be found. We also comment on some
interesting cases including the ones analyzed in the previous
sections.  Finally, in Sec.~\ref{sec-conclusion} we conclude.

\section{Guilfoyle solutions and black hole conditions}
\label{sec-sphericalGuilfoyle0}

\subsection{Spherical static Weyl-Guilfoyle systems and equations}
\label{sec-sphericalGuilfoyle}

The matter 
source is a static charged fluid distribution with spherical symmetry.
The  
metric is conveniently written in  the form
\beq
ds^2 = -B(r)\,dt^2 + A(r)\, dr^2 + r^2\,
(d\theta^2+\sin^2 d\varphi^2)\, ,
\label{metricsph}
\eeq
where $(t,\,r,\, \theta,\, \varphi)$ are spherical
symmetric coordinates, and the functions $A$ and $B$ depend on the radial
coordinate $r$ only.
The fluid is characterized by its energy density $\rho_{\rm m}(r)$, isotropic
pressure $p(r)$, electric charge density $\rho_{\rm e}(r)$, and
four-velocity $ U_\mu =  -\sqrt{B(r)\,}\, \delta_\mu ^t$.
The electromagnetic gauge field ${\cal A}_\mu$ 
assumes the form 
\beq
 {\cal A}_\mu = -\phi(r)\,\delta_\mu^t\,,
\label{electpotentialsph}
\eeq 
where $\phi(r)$ is the electric potential.

Two important auxiliary quantities are the mass $M(r)$
inside a sphere
of radius $r$
\beq
M(r) = \int_0 ^r 4\pi\, r^2 \left( \rho_m(r)
+\frac{Q^2(r)}{8\pi\,r^4}\right)dr + \frac{Q^2(r)}{2\,r} \,,
\label{massdef}
\eeq
and the electric charge $Q(r)$
inside a sphere of radius $r$,
\beq
Q(r) =  4\pi\int_0^r { \rho_{\rm e}(r)\sqrt{A(r)} \,r^2 dr}
\,.
\label{chargedef}
\eeq

The system we study here is a case of the Weyl-Guilfoyle-type
systems for which the metric potential $B(r)$ and the
electric potential $\phi(r)$ are functionally related through a
particular Weyl-Guilfoyle relation \cite{guilfoyle} (see also
\cite{Lemos:2009mr}),
namely, $B(r)= a \left[-\epsilon\,\phi(r)+b\right]^2$,
with $a$ and $b$ being arbitrary constants and $\epsilon=\pm 1$.
Now, the parameter $b$ can be absorbed into 
the electric potential $\phi$, and so without
loss of generality one can set $b=0$. Thus, 
\begin{equation}
B(r)= a \phi^2(r) \, , \label{weylgrel1}
\end{equation}
with $a$ being an arbitrary constant
called the Guilfoyle parameter.

Using the Einstein-Maxwell equations one finds the set of equations
for this system (we use units in which the gravitational constant and
the speed of light are set to one). 
The Einstein part of the equations
furnishes the following relations,
\beqa
&&\hspace*{-1.3cm}\frac{B^\prime(r)}{B(r)}+\frac{A^\prime(r)}{A(r)} =
8\pi r\,A(r)
\Big[\rho_{\rm m}(r) +p(r)\Big]\, ,\label{einsteq1}\\
&&\hspace*{-1.3cm}
 \Big(\frac{r}{A(r)}\Big)^\prime = 1- 8\pi\, r^2\left(
\rho_{\rm m}(r) + \frac{Q^2(r) }{8\pi r^4}\right)\, , \label{einsteq2}
\eeqa
where a prime denotes the derivative with respect to the radial
coordinate $r$.
With the definition of the total charge $Q(r)$
inside a sphere of radius
$r$, see Eq.~(\ref{chargedef}),
the only nonzero component of the
Maxwell equation yields 
$Q(r) =
 \frac{r^{2}\,\phi^\prime (r)}{\sqrt{B(r) \,A(r)}}$,
where an integration constant was set to zero.
From Eq.~\eqref{weylgrel1} we can write $\phi (r)$ in terms of $B(r)$ as
$\epsilon \phi(r) = \sqrt{\dfrac{B(r)}{a}}$.
With this, the amount of electric charge inside a spherical surface of
radius $r$, Eq.~\eqref{chargedef}, is
then given by
\beq
Q(r) = \frac{-\epsilon\,r^{2}\,B^\prime(r)} {\,2\sqrt{a A(r)}\, B(r)}\, .
\label{qspherical2}
\eeq
The electric charge is then obtained once we have the metric functions
$B(r)$ and $A(r)$.

One of the equations,~(\ref{einsteq1}) or (\ref{einsteq2}),
can be interchanged with the contracted Bianchi identity equation,
i.e., the conservation equation given here by
\beq
 2p^\prime(r) + \frac{B^\prime(r)}{B(r)} \Big [\rho_{\rm m}(r) +
p(r)\Big] - 2\frac{\phi^\prime(r) \rho_{\rm e}(r)} {\sqrt{B(r)}}=0\,.
      \label{conserv1}
\eeq
\subsection{Guilfoyle solutions}
\label{sec-solution}

\subsubsection{Interior solution}

The interior goes up to a certain radius $r_0$, say.
Guilfoyle's solutions are found under the assumption that the
effective energy density $\rho_{\rm m}(r) +
\dfrac{Q^2(r)}{8\pi\,r^4}$ is a constant, 
\beq
8\pi\,\rho_{\rm m}(r) + \dfrac{Q^2(r)}{r^4} = \frac{3}{R^2}\, ,
\label{rhoconst}
\eeq
where $R$, a characteristic length associated 
to the inverse of the total energy density, is a constant to 
be related to the parameters of the exterior
solution, $m$ and $q$, by the junction conditions of
the metric at the surface $r=r_0$.
With this hypothesis, Guilfoyle \cite{guilfoyle} 
found an exact solution given by
\beq
A(r) =\left({1 - \dfrac{r^2}{R^2}}\right)^{-1}\, , \label{A(r)1}
\eeq
\beq
B(r)  = \left[\frac{\left(2-a\right)^2}{a^2}\,F^2(r)
\right]^{a/(a-2)} \, ,
\label{B-sol1}
\eeq
\beq
\phi(r) = \epsilon \sqrt{\dfrac{B(r)}{a}}
\label{phi00}
\eeq
\beq
8\pi \rho_{\rm m}(r) = \frac{3}{R^2} -\frac{a}{\left(2-a\right)^2}
\frac{k_0^2}{R^4}\,\frac{r^2} {F^2(r)}\, , 
\label{rhom-sol1} 
\eeq
\beqa
8\pi p(r) &=& -\frac{1}{R^2} + \frac{a}{\left(2-a\right)^2}
\frac{k_0^2}{R^4}\,\frac{r^2} {F^2(r)}      +\nonumber\\ &&
+ \frac{2a}{2-a}\dfrac{k_0}{R^2}\frac{\sqrt{1-r^2/R^2}}{F(r)}\, ,
\label{p-sol1}
\eeqa

\beq
4\pi \rho_{\rm e}(r) = 
\frac{\epsilon \sqrt{a}}{2-a}\frac{k_0^2}{R^4}\frac{r^2}{F^2(r)}\!
\left(\!1+
\frac{3F(r)\sqrt{1-r^2/R^2}}{k_0r^2}\!\right).
\label{rhoe}
\eeq
The auxiliary functions are
\beq
M(r) =
\frac{r^3}{2R^2}+\frac{a}{2\left(2-a\right)^2}\frac{k_0^2}{R^4}\frac{r^5}{
F^2(r) },
\label{mass-funct}
\eeq
\beq
Q(r) = \frac{\epsilon \sqrt{a\,}}{2-a}\frac{ k_0}{R^2}\,
\frac{r^3}{F(r)}\,.
\label{charge-sol1}
\eeq
The function $F(r)$ is defined by
\beq
F(r) =  k_0\, \sqrt{1 - \frac{r^2}{R^2}}-k_1 \, ,\label{Fr}\\
\eeq
and the integration constants $k_0$ and $k_1$ are
given by
\beq
k_0 =\frac{R^2}{r_0^2}
\left(\frac{m}{r_0}-\frac{q^2}{r_0^2}\right) \left(1- \dfrac{r_0^2}
{R^2}\right)^{-1/a}\, ,
\label{constk}
\eeq
\beqa
k_1&=& \!\!k_0\,\sqrt{1-\dfrac{r_0^2}{R^2}}\!
\left[1 - \frac{a}{2-a}\!\dfrac{r_0^2}{R^2}
\left(\dfrac{m}{r_0}-\dfrac{q^2}{r_0^2}
\right)^{\!\!-1}\right], 
\label{constk_1}
\eeqa
where
\beq
m\equiv M(r_0)\,,
\label{m_1}
\eeq
\beq
q\equiv Q(r_0)\,.
\label{q_1}
\eeq
The constants $k_0$ and $k_1$ are found through the
junction conditions.
As one can check, the above solution is valid for all $a>0$, the limiting
case $a \rightarrow\infty$ yielding the uncharged ($q=0$) Schwarzschild
interior solution.
We are not interested in negative $a$.

\subsubsection{Exterior solution}

The solution of Einstein-Maxwell field equations, 
Eqs.~\eqref{einsteq1}-\eqref{qspherical2}, for the external region, $r>
r_0$, is given by
\beq
{A(r)} =\frac{1}{1 -\dfrac{2m}{r}+ \dfrac{q^2}{r^2}}\,,
\label{ABext1}
\eeq
\beq
B(r) = \dfrac{1}{A(r)} = 1 -\dfrac{2m}{r}+ \dfrac{q^2}{r^2}\,,
\label{ABext2}
\eeq
\beq
\phi(r) = \dfrac{q}{r}\,, 
\label{phiext}
\eeq
\beq
\rho_{\rm m}(r)=0\,, 
\label{fluidext1}
\eeq
\beq
p(r)=0\,, 
\label{fluidext2}
\eeq
\beq
\rho_{\rm e}=0\,, 
\label{rhoeext}
\eeq
and the auxiliary functions are
\beq
M(r) = m \,,  
\label{massext}
\eeq 
\beq
Q(r) = q \,,  
\label{chargeext}
\eeq
which is the Reissner-Nordstr\"om solution. Note that, by continuity
on the surface $r=r_0$, the metric functions yield $B(r_0)= 1/A(r_0)=
1 -\dfrac{2m}{r_0}+ \dfrac{q^2}{r_0^{2}}$. Also one must have
$\phi(r_0) = \dfrac{q}{r_0}$, $M(r_0)=m$, and $Q(r_0) =q$,
$m$ and $q$ being the total mass and total charge of the
exterior spacetime.

\subsubsection{Junction conditions}

To do the matching properly, we start by imposing the
junction condition that the metric should be continuous 
at the boundary surface $r=r_0$. In fact, by joining the interior metric
function $g_{rr}=A(r)$ in Eq.~\eqref{A(r)1} with the 
$g_{rr}$ coefficient of
the exterior metric given by \eqref{ABext2}, it is found
\cite{guilfoyle} that
\beq
 m= \frac{r_0}{2}\left(\frac{r_0^2}{R^2} +
\frac{q^2}{r_0^2}\right).
\label{mass}
\eeq
Another junction condition arises by
using the continuity of the $g_{tt}=B(r)$ metric coefficient,
the continuity of its first derivative at $r=r_0$, and
Eq.~\eqref{qspherical2}. Then one gets 
\beq
a=
\frac{r_0^2}{4q^2}\left(\frac{r_0^2}{R^2} -\frac{q^2}{r_0^2}\right)^2
\left(1-\frac{r_0^2}{R^2}\right)^{-1}. \label{a-function}
\eeq
where the fact that $Q(r_0)= q$ was also taken into account, and positive
$q$ is assumed without loss of generality.

\subsubsection{Some constraints}
\label{sec-someconstraints}

Equations~\eqref{mass}-\eqref{a-function} are two constraints to the
five free parameters of the solutions, $a$, $R$, $m$, $q$, and $r_0$. 
We now investigate some constraints on these parameters.

Using Eqs.~\eqref{mass}-\eqref{a-function} it is possible to
treat  the Guilfoyle parameter $a$ and the
mass $m$ of the solution as a function of
the other parameters of the theory, namely, $m=(r_0,R,q)$ and $a=a(r_0,
R,q)$.  Relation \eqref{a-function}
implies $a$ may be zero or negative in some range of the parameters.
However, to avoid imaginary electromagnetic fields, 
we are interested just in solutions for which 
\begin{equation}
a>0
\,,
\label{a>0}
\end{equation}
and so we use Eq.~\eqref{a-function} to constrain the range of the
parameters $r_0$ and $R$. The condition $a>0$ requires that
\begin{equation}
{r_0} \leq {R} \,. \label{r0 R}
\end{equation}

The constants $k_0$ and 
$k_1$ in Eqs.~(\ref{constk})-(\ref{constk_1}) 
are also found through the junction conditions.

\subsection{Black hole conditions}
\label{sec-blackholeconditions}

We are looking for regular black holes. The conditions for the solutions to
be regular are subtle and can be found numerically. The conditions for the
solutions to be black holes are simple and can be stated now. The first
condition to check is the presence of horizons. Namely, the expressions
for  the Cauchy and event horizon radii of the Reissner-Nordstr\"om solution
exterior to the matter, 
\beq
r_- = m - \sqrt{m^2- q^2}\, , \label{rpm1}
\eeq
\beq
r_+ = m + \sqrt{m^2- q^2}\, , \label{rpm}
\eeq
respectively,
have to produce real numbers. Then the overcharged solutions are discarded
and we have the constraint,
\begin{equation}
 m^2\geq q^2 \,. \label{m>=q}
\end{equation}
Thus the existence of $r_-$ and $r_+$ for the electrically charged solutions
guarantees that the solutions are black holes.

Moreover, the radius of the boundary surface $r_0$ cannot be larger
than, or at most, the gravitational radius $r_+$, i.e., $r_0 \leq r_+$. 
However, inequality \eqref{m>=q} together with  condition 
\eqref{r0 R} imply that the value of $r_0$ cannot be in the region
between $r_-$ and $r_+$. Additionally, the boundary conditions impose that
$r_0$ cannot be equal to $r_+$, with one special case as an exception, when
$r_0=r_-$ and at the extremal regime $m^2=q^2$, which means $r_-=r_+=r_0$. 
The constraint for $r_0$ is then 
\beq
0\leq r_0 \leq r_-\,,   \label{r0-rminus}
\eeq
the equality holding in the limit of $r_0=R$.
This result and Eq.~\eqref{rpm} give
\beq
0\leq r_0 \leq m \, , \label{r0-mq}
\eeq
the equality holding in the limit of $q=m$.

\section{Regular nonextremal black holes with charged phantom matter}
\label{sec-rbhgen}

In testing for regularity of the solutions, we find that 
there is a region of the parameter space which corresponds to regular black
holes for small characteristic length $R$, namely,
\begin{equation}
0 <
\frac{R}{q}\leq 1
\,.
\label{Rless1}
\end{equation}
From now on we write the parameters of the problem as 
parameters without units, mainly we use the total charge $q$ 
as the quantity to which the other parameters are gauged. Since from 
Eq.~(\ref{rhoconst}), $\rho_{\rm m}(r) +
Q^2(r)/8\pi r^4 = 3/8\pi R^3={\rm constant}$, small characteristic
length $R$ means
sufficiently large total energy density, i.e., the configurations are
compact relative to the electric charge. Or, if one prefers,
the configurations are heavily charged in comparison.

\begin{figure}[pb]
\centerline{\includegraphics[scale=1.1]{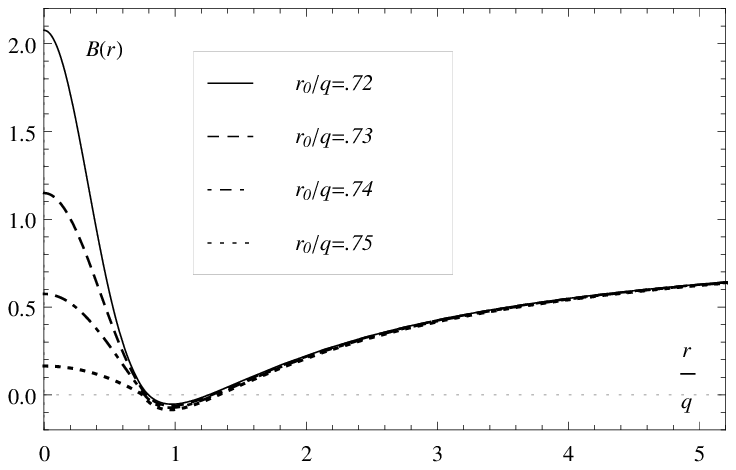}}
\caption{Graphs of $B(r)$ for
$R/q=0.75$ and four different ${r_0}/{q}$: ${r_0}/{q}
= 0.72$ (solid line), $r_0/q= 0.73$ (dashed line),
$r_0/q = 0.74$ (dash-dotted line), and $r_0/q = 0.75$ (dotted line).}
\label{plot-B3}
\vskip 0.5cm
\centerline{\includegraphics[scale=1.1]{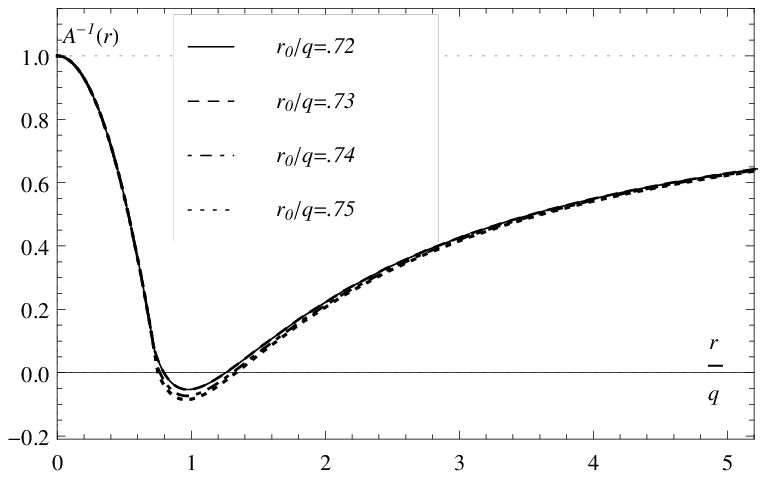}}
\caption{Plots of $1/A(r)$ for $R/q=0.75$ and four different 
${r_0}/{q}$: $r_0/q = 0.72$ (solid line), $r_0/q= 0.73$ (dashed line),
$r_0/q = 0.74$ (dash-dotted line), and $r_0/q = 0.75$ (dotted line).}
\label{plot-A3}
\end{figure}

Besides $R/q$, another useful parameter for
our analysis is the ratio $r_0/R$, which is the ratio of the radius of
the matter distribution to the energy density parameter $R$.  For very
compacted charged spheres, typically, for $r_0/R \lesssim 0.9$, the fluid
energy density $\rho_{\rm m}(r)$ assumes negative values at some
$r$. On the other hand, for 
\begin{equation}
\frac{r_0}{R}\simeq 1 \label{r0overR1}
\end{equation}
from below, the energy density $\rho_{\rm m}(r)$ is positive, while the 
pressure is negative everywhere inside the central core.
For solutions in this range one has $p(r)< -\rho_{\rm m}(r)$, i.e., 
\begin{equation}
\rho_{\rm m}(r)+p(r)<0\,, \label{prho-1}
\end{equation}
which means that the 
fluid is composed of a  phantom matter with electric charge.
This is a Guilfoyle phantom fluid.

To show that these configurations are indeed black holes, 
regular ones, 
we focus on the physical quantities $B(r)$, $A(r)$,
$\rho_{\rm m}(r)$, $p(r)$, and $\rho_{\rm e}(r)$ and plot
them as functions of $r$.
The potential $\phi(r)$ is not necessary to plot as it is
given by Eq.~(\ref{phi00}). We also plot $M(r)$ and 
$Q(r)$ which are useful quantities.
As a typical class of configurations we consider the case
\begin{equation}
\frac{R}{q}=0.75 \,. \label{Rq0.75}
\end{equation}
According to Eq.~\eqref{r0 R}, the maximum value for $r_0/q$ is 
$r_0/q=R/q$, so in this case the upper limit is then 
\begin{equation}
\frac{r_0}{q}\leq0.75 \,. \label{R0q0.75}
\end{equation}

\begin{figure}[ph]
\centerline{\includegraphics[scale=1.1]{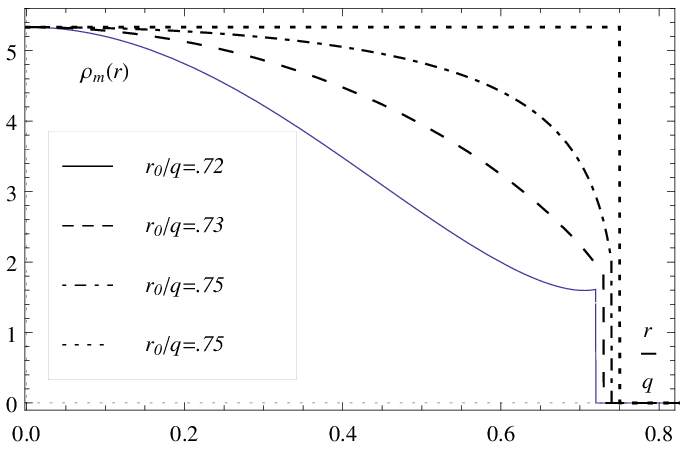} }
\caption{A plot of the energy density $\rho_{\rm m}(r)$ for 
$R/q=0.75$ and
four
different ${r_0}/{q}$: $r_0/q
= 0.72$ (solid line), $r_0/q= 0.73$ (dashed line),
$r_0/q = 0.74$ (dash-dotted line), and $r_0/q = 0.75$ (dotted line).}
\label{plot-Rho1}
\vskip 0.5cm
\centerline{\includegraphics[scale=1.]{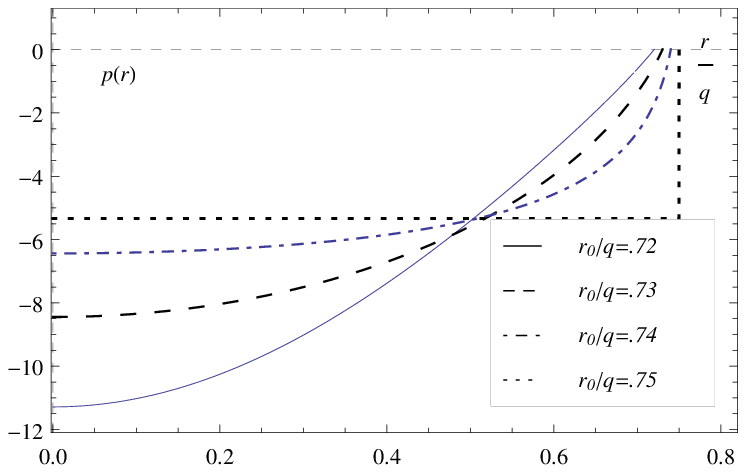} }
\caption{A plot of the pressure $p(r)$ for $R/q=0.75$ and four
different ${r_0}/{q}$: $r_0/q
= 0.72$ (solid line), $r_0/q= 0.73$ (dashed line),
$r_0/q = 0.74$ (dash-dotted line), and $r_0/q = 0.75$ (dotted line).}
\label{plot-p1}
\end{figure}

\begin{figure}[pb]
%\vskip 0.5cm
\centerline{\includegraphics[scale=1.15]{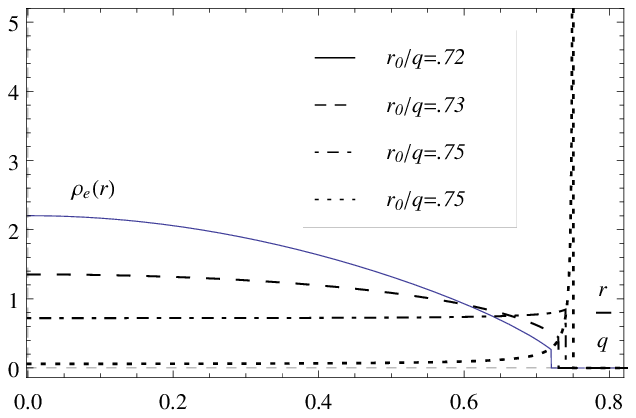} }
\caption{A plot of the charge density $\rho_{\rm e}(r)$ for $R/q=0.75$
and four different ${r_0}/{q}$: $r_0/q = 0.72$ (solid line), $r_0/q=
0.73$ (dashed line), $r_0/q = 0.74$ (dash-dotted line), and $r_0/q =
0.75$ (dotted line).} \label{plot-Rho_e1}
\end{figure}

\begin{figure}
\centerline{\includegraphics[scale=1.2]{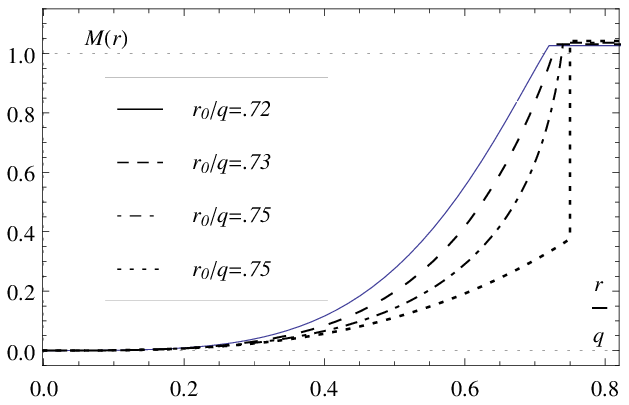} }
\caption{A plot of the mass function $M(r)$ for $R/q=0.75$ and four
different ${r_0}/{q}$: $r_0/q
= 0.72$ (solid line), $r_0/q= 0.73$ (dashed line),
$r_0/q = 0.74$ (dash-dotted line), and $r_0/q = 0.75$ (dotted line).}
\label{plot-mass-funct}
\end{figure}

\begin{figure}
\centerline{\includegraphics[scale=1.2]{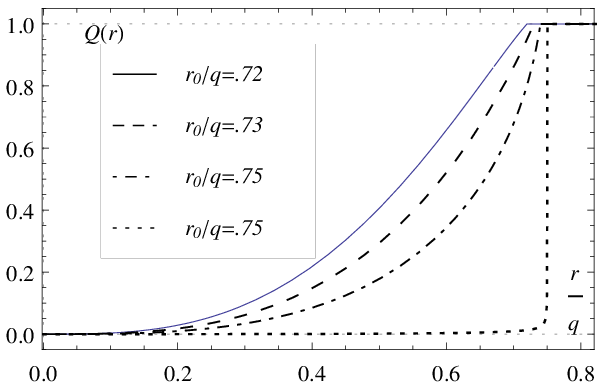} }
\caption{A plot of the charge function $Q(r)$ for $R/q=0.75$ and four
different ${r_0}/{q}$: $r_0/q
= 0.72$ (solid line), $r_0/q= 0.73$ (dashed line),
$r_0/q = 0.74$ (dash-dotted line), and $r_0/q = 0.75$ (dotted line).}
\label{plot-Qe}
\end{figure}

In Figs.~\ref{plot-B3}-\ref{plot-Qe} we display the behavior 
as a function of $r$ of the
above-mentioned physical quantities for four different values of
$r_0/q$, namely, ${r_0}/{q} = 0.72,\, 0.73,\,0.74,\,0.75$, this latter
number being the maximum value possible, see Eq.~(\ref{R0q0.75}).

In Figs.~\ref{plot-B3}-\ref{plot-A3} the plots of the metric
potentials $B(r)$ and $1/A(r)$ are shown. Clearly there are no curvature
singularities, i.e., the black holes are regular. The curve for $r_0/q
= 0.75$ (dotted line) corresponds to a regular black hole with a de
Sitter core, to be
treated separately.  The small difference among the curves
of Figs.~\ref{plot-B3} and~\ref{plot-A3} is related to the fact that
we considered regular black holes of almost the same sizes, i.e., with
values of $r_0/q$ very close to each other.  Although the values of
the mass $m$ are relatively close to $q$, these regular black holes
are far from being extremal as $r_+$ and $r_-$ differ substantially.
These facts can be found in Table I, which shows the values of the mass
$m$, the Cauchy and event horizon radii, $r_-$ and $r_+$,
respectively, for some chosen values of $r_0/q$, with $R/q=0.75$
fixed.  It is clear from the table that $r_0<r_-$, i.e., the matter
radius is inside the Cauchy horizon.

\begin{table}[h]
\begin{center}
\begin{tabular}{|c|c|c|c|}
\hline
$r_0/q$   &\; $m(r_0,0.75)/q\;$ &\; $r_-(r_0,0.75)/q$\; &\; 
$r_+(r_0,0.75)/q$\;
\\
\hline
  0.720000  &   1.02622  &  0.795724  &  1.25672\\
  0.730000  &   1.03072  &  0.780939  &  1.28051\\
  0.740000  &   1.03587  &  0.765622  &  1.30613\\
  0.750000  &   1.04167  &  0.750000  &  1.33333\\
       \hline
\end{tabular} \label{table1}
\caption{Values of the mass $m$, Cauchy horizon $r_-$ and event
horizon $r_+$ in terms of the radius $r_0$ of the charged mass core of
the regular black holes for $R/q=0.75$, for 
four different values of $r_0/q$.}\end{center}
\end{table}


\begin{figure}[h]
\begin{center}
\includegraphics[scale=.42]{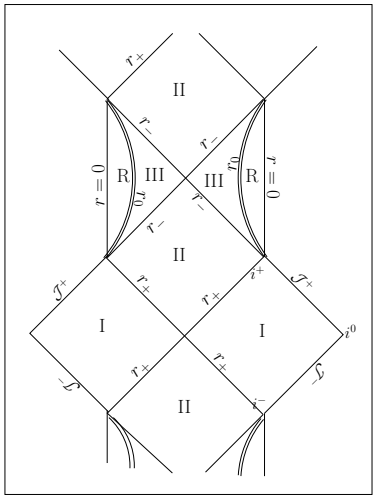}
\caption{The Carter-Penrose diagram for the regular nonextremal 
black
hole with $r_0<r_-< r_+$. }
\label{pen-timelk}
\end{center}
\end{figure}

In Fig.~\ref{plot-Rho1} the energy density $\rho_{\rm m}(r)$ is displayed 
for the mentioned range of parameters. We see that within this range it is 
positive. The curve with $r_0/q = 0.75$ (dotted line) is flat, has 
a sharp drop at the boundary, and corresponds to a de Sitter core. Due
to its particular interest it is
studied separately. For smaller values of $r_0$
the solutions are 
less interesting, as they have negative energy densities. In 
Fig.~\ref{plot-p1} the pressure $p(r)$ is shown. The values are negative and 
higher in modulus than the  energy density $\rho_{\rm m}(r)$. The fluid is a 
phantom fluid. The curve with $r_0/q = 0.75$ (dotted line), that is flat and 
drops sharply to zero at the boundary corresponds to a de Sitter core. In 
Figs.~\ref{plot-Rho_e1} the charge density profiles $\rho_{\rm m}(r)$
for the given range of
parameters are shown. Finally, the behavior of the auxiliary functions
$M(r)$ and $Q(r)$ is shown in Figs.~\ref{plot-mass-funct} and
\ref{plot-Qe}, respectively.

The conformal Carter-Penrose 
diagram of this regular black hole is shown in
Fig.~\ref{pen-timelk}. The external region ($r>r_0$) is
Reissner-Nordstr\"om, possessing a Cauchy and an event horizon, $r_-$
and $r_+$, respectively. In the external region there are three
distinct regions indicated in the figure by I, II, and III. The
internal regular region, indicated by R in the diagram, is filled
by a distribution of charged matter for which the pressure is negative
and larger in absolute value
than the energy density in the central core.

\newpage

\section{Regular black holes with a de Sitter core and a massless
charged layer at the Cauchy horizon}
\label{sec-regdeSitterC}

Within this set of Guilfoyle solutions 
there are also regular black holes with a de Sitter core 
whose matching to the exterior region occurs at
the Cauchy horizon,  $r_0=r_-$, which is a null, i.e., lightlike, 
surface. 
The interior fluid, $r<r_0=r_-$,  is
uncharged, satisfying a de Sitter equation of state $p=- \rho_m =
3/8\pi R^2$, so that 
\begin{equation}
\rho_{\rm  m}(r)+p(r)=0\,,
\label{prho=0}
\end{equation}
for this fluid. It is a false vacuum de Sitter core, the limiting case of a
phantom fluid $\rho_m +p<0$.
The boundary of this de Sitter core, $r=r_0=r_-$, is a layer of zero mass
but nonzero charge.
Thus, all the electric charge is on such a surface.

Let us show from the equations that this regular black 
hole solution exists.

By inspection, or otherwise, it is seen that this solution exists when 
$r_0=r_-$. Since by 
Eq.~(\ref{rpm1}), $r_-= m -\sqrt{m^2-q^2}$, using Eq.~(\ref{mass}),
we obtain 
$r_0= m(q,r_0,R)-\sqrt{m^2(q,r_0,R)-q^2}
  ={q^2}/{2r_0} + {r_0^3}/{2R^2} -
\sqrt{\left({q^2}/{2r_0} +{r_0^3}/{2R^2}\right)^2 -q^2}$.
%\begin{eqnarray}
% r_0&=& m(q,r_0,R)-\sqrt{m^2(q,r_0,R)-q^2} \nonumber\\
%  &=& \frac{q^2}{2r_0} +\frac{r_0^3}{2R^2} -
%\sqrt{\left(\frac{q^2}{2r_0} +\frac{r_0^3}{2R^2}\right)^2 -q^2}\,.
%\end{eqnarray}
This equation gives two solutions. One is $r_0=r_- = 0$, which is satisfied
for all $q^2\geq 0$ and is in fact a singular black hole.
The other solution is
\begin{equation}
r_0=r_- = R \,,
\label{r0r-R}
\end{equation}
and is satisfied only for ${R}/{q}<1$. This latter is the solution that
yields regular black holes. Thus 
\begin{equation}
\frac{r_0}{q}=\frac{r_-}{q}=\frac{R}{q}<1\,,
\label{allroverqless1}
\end{equation}
for this black hole.

Let us now discuss the features of the three
different regions of this regular black hole.

\vskip 0.2cm
\noindent
(i) The region $r<r_0=r_-=R$:
In this region the interior fluid is
uncharged, satisfying a de Sitter equation of state. 

Indeed, in the limit $r_0 \rightarrow R$ the ratio $-p(r)/\rho_{\rm
m}$ tends to unity, $p=- \rho_m = 3/8\pi R^2$.
One can find this result by taking the limit $r_0\rightarrow R$ in
Eqs.~\eqref{B-sol1}-\eqref{constk_1}
with the help of Eqs.~\eqref{mass}-\eqref{a-function}.

To be specific, 
we see from Eq.~\eqref{a-function} that for  $r_0=R$ and for
${r_0}/{q}<1$, see Eq.~\eqref{allroverqless1}, the parameter $a(r_0,R)$
diverges with $\displaystyle{(1-r_0^2/R^2)^{-1}}$.  Hence,
Eq.~\eqref{constk} implies that the following limit,
$\displaystyle{\lim_{r_0\rightarrow R} k_0(r_0,R)}= \frac{R^2}{r_0^2}
\left(\frac{m}{r_0}-\frac{q^2}{r_0^2}\right)$, is finite. Similarly, it
follows from Eq.~\eqref{constk_1} that the constant $k_1$ goes to zero with
$\displaystyle{\sqrt{1-r_0^2/R^2}}$.
Therefore, for $r\neq R$, the function $F(r)$ defined in Eq.~\eqref{Fr} is
finite and nonzero. Then, in the limit ${r_0\rightarrow R}$ the solution for
$r< r_0$ is
\beqa
 & & B(r)  = \dfrac{1}{4}\left(1-\frac{q^2}{r_0^2}\right)^2\left( 1 -
\dfrac{r^2}{R^2}\right)\nonumber \\ \nonumber\\ && =
\dfrac{1}{4}\left(1-\frac{q^2}{r_0^2}\right)^2A^{-1}(r)\, ,
 \quad r< r_0\,, \label{B-sol2}\\
 & & 8\pi\, \rho_{\rm m}(r) =  -8\pi\, p(r)=\frac{3}{R^2} \,, \quad r< r_0\,,
\label{rhom-p-sol2} \\
& & Q(r) = 0 \,, \quad r< r_0\,  . \label{charge-sol2}
\eeqa
Thus these black holes have a central de Sitter
vacuum up to $r_0$, $r<r_0$.

Note that generically, the metric function $B(r)$ can be made equal to
$1/A(r)$ by a redefinition of the time coordinate. Note also that for
$r_0/q=1$, a case that follows outside the configurations we are studying
(see Eq.~\eqref{allroverqless1}), $B(r)=0$ for all $r<r_0$ and we are
in the presence of a quasiblack hole.

\begin{figure}[h]
\begin{center}
\includegraphics[scale=.42]{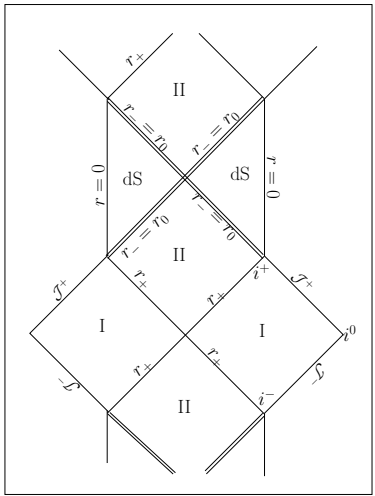}
\caption{The Carter-Penrose diagram for the regular black
hole with a de Sitter central core and a lightlike matter boundary
at the Cauchy horizon, where $r_0<r_-$.}
\label{pen-null}
\end{center}
\end{figure}

\vskip 0.2cm
\noindent
(ii) The region $r=r_0=r_-=R$: In this region one has an electrically
charged massless layer. Indeed, in the limit $r_0 \rightarrow R$,
and $r_0/q< 1$ the function $F(r)$ defined in Eq.~\eqref{Fr}
goes to zero as $\displaystyle{\sqrt{1-r_0^2/R^2}}$.
Then the charge density diverges at $r=r_0=R$, but the total charge is
finite, 
\begin{equation}
Q(r=r_0)=q\,, \quad r=r_0 \,.
\label{rr-R}
\end{equation}

\vskip 0.2cm
\noindent
(iii) The region $r>r_0=r_-=R$: 
By construction this region is the Reissner-Nordstr\"om vacuum.

Thus, we have shown that a regular black hole solution
with a de Sitter core and a massless electrically
charged layer at the Cauchy horizon is a solution of
the equations.

The resulting metric and fluid functions for this extremal case
correspond to the dotted curves of Figs.~\ref{plot-B3}-\ref{plot-Qe}. 
Those plots show the particular case $r_0/q=R/q= 0.75$.
Worth noting is the graph for
$\rho_{\rm e}(r)$ (dotted curve of Fig.~\ref{plot-Rho_e1}), which is
essentially $0$ for $0\leq r< r_0$, and diverges at $r=r_0$. Its
integration gives $Q(r)$, see Fig.~\ref{plot-Qe}, 
i.e., it is zero for $r< r_0$ and $q$ for $r\geq r_0$,
indicating that $\rho_{\rm e}$ has similar properties to the Dirac
delta function.
This special case is  identical to a special case of the regular charged
black holes found in \cite{Lemos:2011vz}, see also \cite{barisrl91}. 
The conformal Carter-Penrose diagram for these regular black holes is shown
in Fig.~\ref{pen-null}. 
In the external region there are two
distinct regions indicated in the figure by I  and II. The
internal regular region, indicated by dS in the diagram, the de Sitter core,
is fulfilled by an uncharged fluid with equation of state of de Sitter type,
$p=-\rho_m$.

\section{The space of parameters: looking for all regular black holes}
\label{sec-analysis}

\subsection{Generics}

The model has five free parameters, $m$, $q$, $r_0$, $R$, and $a$, and 
two relations among them (Eqs.~\eqref{mass} and \eqref{a-function}). In
summary, we have five parameters and only two constraints so that three
parameters are left free. Here we choose $r_0$, $R$, and
$q$ as free parameters. 
We normalize all quantities in terms of $q$, which is
equivalent to setting the charge $q$ to unity.

In the two previous sections we studied two typical classes
of regular black holes within the Guilfoyle solutions. Now, we 
investigate the space of parameters where regular black holes can be found.
For this, we start by showing the dependence of the Guilfoyle parameter $a$
and of the mass $m$ as a function of the other parameters of the model.
After that we show all the different regions containing regular black holes.

\subsection{The mass $m$ and the Guilfoyle parameter $a$}
\label{sec-mass_a}

\subsubsection{The mass function $m$}

\begin{figure}
\centerline{\includegraphics[scale=1.15]{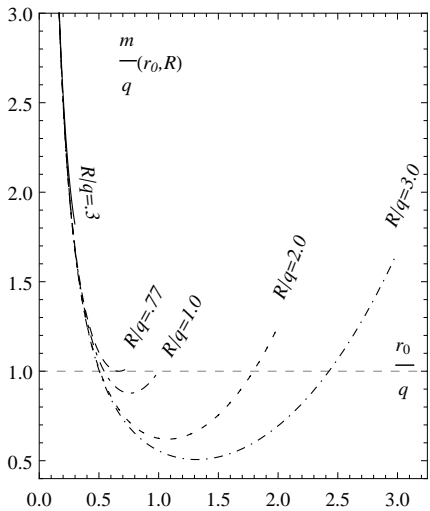}}
\caption{A plot of the mass $m/q$ as a function of $
r_0/q$ for different values of $R/q$, as shown in the curves.}
\label{plot-mass}
\end{figure}

From Eq.~(\ref{mass}) it follows that the ratio $m/q$ is a function of the
two parameters $r_0/q$ and $R/q$.
Figure~\ref{plot-mass} shows the mass $m/q$ of each configuration 
as a function of $r_0/q$ for five different values  of $R/q$.  Since $r_0$ 
cannot be larger than $R$, all the plotted curves in 
such a figure end at $r_0=R$.

Based on Eq.~(\ref{mass}) and Fig.~\ref{plot-mass} we find that there is
an important value for $R/q$, namely, ${R}/{q} =4\sqrt{3}/9$, with
$4\sqrt{3}/9\simeq 0.77$. We can then divide the analysis of the mass
properties $m/q$ into three cases depending on the values of $R/q$: (i) 
$0<{R}/{q} <{4\sqrt{3}}/{9}$, (ii) ${R}/{q} =4\sqrt{3}/9$,
and  (iii) $4\sqrt{3}/9<{R}/{q} <\infty$.

\vskip 0.2cm
\noindent
(i) $0<{R}/{q} <{4\sqrt{3}}/{9}$: In this region there appears also
another important value for $R/q$, ${R}/{q} =\sqrt{3}/3$, with
$\sqrt{3}/3\simeq 0.56$. (a) 
For $0<{R}/{q} <\sqrt{3}/3$ all of the curves
end abruptly when $a=0$
(the region of the
parameter space where the Guilfoyle parameter $a$ assumes negative
values is not considered in the present work as for these
values the electric charge is imaginary). This happens, e.g., in the
case of the curve for $R/q= 0.3$, see Fig.~\ref{plot-mass}, 
which ends at an $r_0$ 
given by $r_0/q =0.3$.  
(b) For ${R}/{q}
=\sqrt{3}/3$ one finds the curve for which the minimum coincides
exactly with the endpoint of the line.  This curve is not shown in
Fig.~\ref{plot-mass}, because it would cause a mess in the figure.
(c) 
For $\sqrt{3}/3<{R}/{q} <4\sqrt{3}/9$ and for fixed $R/q$ the mass
function $m(r_0,R)$ has a minimum at some finite value of $r_0$, see
Fig.~\ref{plot-mass}. This minimum is
denoted by $\bar r_0$.  Now,
Eq.~\eqref{mass} furnishes for this minimum the value $\bar r_0/q=
\sqrt[4]{R^2/3q^2\,}$.  The corresponding minimum value of the mass,
$\bar m = m(\bar r_0,R)$, is such that $\bar m \bar r_0 = 2q^2/3$.
Thus, in brief, within this range, for fixed $R/q$, we see that as
$r_0/q$ increases the mass $m/q$ increases if $r_0/q >
\sqrt[4]{R^2/3q^2}$ and the mass $m/q$ decreases if $r_0/q <
\sqrt[4]{R^2/3q^2}$.  The increasing of the mass $m$ with $r_0$ for
large $r_0$ is expected, because with $R/q$ fixed, i.e., with the
energy density fixed, the mass-energy increases with the size of the
object.  On the other hand, the decreasing of $m$ with $r_0$ for small
$r_0$ may be understood by noticing that in this case the total mass
is dominated by the electromagnetic energy, $q^2/2r_0$, which
decreases with increasing $r_0$.

\vskip 0.2cm
\noindent
(ii) ${R}/{q} ={4\sqrt{3}}/{9}$: For this value the minimum of the curve
$m/q$ coincides with the value $m/q=1$. This minimum
limit is found at $\bar r_0/q=2/3$, see Fig.~\ref{plot-mass}.

\vskip 0.2cm
\noindent
(iii) ${4\sqrt{3}}/{9}<{R}/{q} <\infty$: 
In this case the value of $\bar r_0$ is larger than $\bar r_0/q \simeq 0.42$
and the respective masses are such that $ m/q \leq 1$. In fact there is a
range of values of $r_0$ for which $m/q \leq1$, meaning these objects are
overcharged or are extremal. Overcharged objects certainly do not represent
regular black holes and are not of interest for the present work. Extremal
black holes that are regular could not be found in these solutions. Thus,
only configurations that satisfy $m/q>1$ can be regular black holes. Within
this range of $R/q$ we see that regular black holes can exist in the two
disjoint parts of the curve for which $m/q\geq1$. 
An example is $R/q=2$. It
has $\bar r_0/q = \sqrt[4]{4/3\,}$, and $\bar m/q = \sqrt[4]{12\,}/3< 1$, 
see Fig.~\ref{plot-mass}.

\subsubsection{The Guilfoyle parameter $a$}

\begin{figure}
\centerline{\includegraphics[scale=1.15]{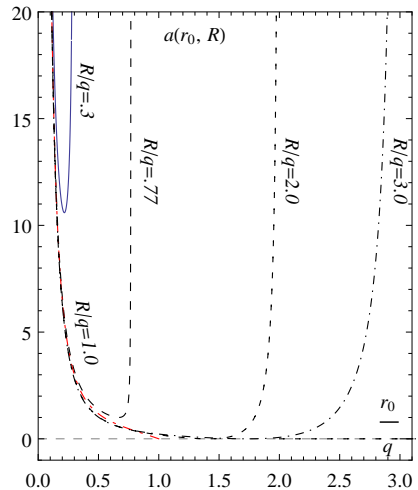}}
\caption{A plot of the parameter $a$ as a function of $r_0/q$ 
for different values of $R/q$. All the curves
have two branches, one for positive values and the other for negative values.
The branches of negative values are not shown. The function $a(r_0/q,R/q)$ is
singular at $r_0/q=R/q$. The only exception is for $R/q=1$, which gives
$a(r_0/q,R/q)=0$.}
\label{plot-a}
\end{figure}

Typical curves for the parameter $a$ as a function of $r_0/q$ and $R/q$ are
shown in Fig.~\ref{plot-a}. For fixed $R/q$ one has that all the curves show
a smooth branch in the form of a cup. The curves have a minimum at some
value of $r_0$, $\tilde r_0$ say (the root of a polynomial), and tend to
$+\infty$ at  $r_0 \rightarrow 0$ and $r_0\rightarrow R$. Examples of these
are the curves for $R/q = 0.3$, $R/q = 0.77$, $R/q = 2.0$, and $R/q = 3.0$
in Fig.~\ref{plot-a}. The case $R/q=1$ is special in the sense that the
curve has no local minimum and ends at ${r_0}/{q}=1$.

\subsection{Regions of the parameter space and general properties
of the solutions}
\label{sec-regions}

In the following we list the different regions of the parameter
space $(r_0/q,R/q)$ in which the corresponding solutions represent
regular black holes.

\subsubsection{$0\leq R/q < 4\sqrt{3}/9$: High mass-energy densities}

\begin{figure}[h]
\centerline{\includegraphics[scale=1.]{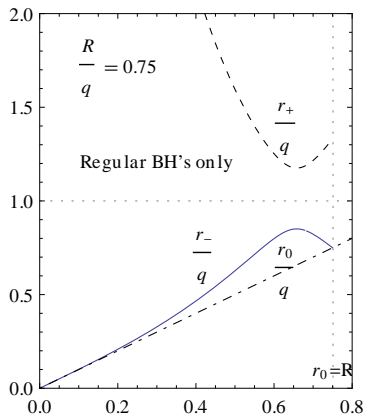}}
\caption{A plot of $r_+/q$ and $r_-/q$ as a function of $r_0/q$ 
for $R/q=0.75$, the case explored in Sec.~\ref{sec-analysis}}.
\label{plots-rs.75}
\end{figure}

For
sufficiently compact charged mass distributions, i.e., high mass-energy
densities, $R/q$ is in the range $0\leq R/q < 4\sqrt{3}/9\simeq 0.77$. A
typical case in this range is shown in Fig.~\ref{plots-rs.75} for $R/q=
0.75$.
Particular instances of $r_0/q$ within this
case were studied in Sec.~\ref{sec-rbhgen}.

In Fig.~\ref{plots-rs.75} we plot the horizon radius $r_+$ and the Cauchy
horizon $r_-$ in terms of $r_0/q$ for $R/q= 0.75$.
There are regular black holes for all values of $r_0/q$. 

The dependence of
the mass $m$ and of the parameter $a$ in terms of $r_0/q$ 
for this $R/q= 0.75$ case is similar to
the curves indicated by $R/q = 0.3$ and  $R/q = 0.5$ in
Figs.~\ref{plot-mass} and \ref{plot-a}, respectively.
One notes that the mass is larger than
the charge for all these objects. 
Also for these large mass-energy density objects
there are always two horizons, the event horizon $r_+$ and the Cauchy
inner horizon $r_-$ which cannot be equal to each other, $r_+(r_0,R)
> r_-(r_0,R)$. 
Thus there are no extremal black holes.
The boundary of the matter core $r_0$ can be equal to the
Cauchy horizon only when $r_0 = R$ ($r_0 =0$ is excluded
because it gives singular solutions). Solutions belonging to this region
of the parameter space may be regular black holes with a timelike
boundary below the inner
horizon, or, if $r_0=R$, with a lightlike boundary at the inner
horizon, see Secs.~\ref{sec-rbhgen}-\ref{sec-regdeSitterC}.

The good regular black holes, i.e., the ones with
non-negative fluid energy density, are found for $r_0/q$ close to $R/q$. Such
a kind of situation is shown by the particular solutions whose metric,
energy density
pressure, and charge density functions are plotted in
Figs.~\ref{plot-B3}-\ref{plot-Qe}.

The solutions for small $r_0$ tend to behave badly in the sense that the
pressure $p(r)$ assumes very large negative values in some regions. In 
the same
situations, the energy density of the fluid $\rho_{\rm m}(r)$ may also become
negative in some regions. 

\subsubsection{$R/q =4\sqrt{3}/9$: Special case for
the mass-energy density}

The dependence of the mass $m$ in terms of $r_0/q$ for this $R/q
=4\sqrt{3}/9\simeq 0.77$ case is indicated in Fig.~\ref{plot-mass}, see also
Fig.~\ref{plot-a} for the dependence of $a$. The mass is larger than the
charge for all these objects, except at the extremal value, 
$r_0/q=2/3$, for which the mass equals the electric charge and the two
horizon radii are identical, $r_+/q=r_-/q = 1.0$.

\begin{figure}[h]
\centerline{\includegraphics[scale=1.]{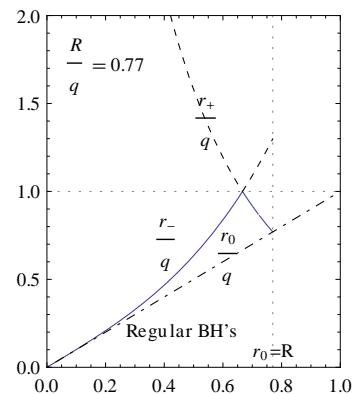}}
\caption{A plot of $r_+/q$ and $r_-/q$ as a function of $r_0/q$ 
for  $R/q =4\sqrt{3}/9\simeq 0.77$. The horizons coincide at
$r_0/q=2/3$ for which $r_+/q=r_-/q=1$.}
\label{plots-rs.77}
\end{figure}

The plots of the radii $r_\pm/q$ in terms of $r_0/q$ for this special
value of $R/q$ are shown in Fig.~\ref{plots-rs.77}.  For $r_0/q\neq 2/3$, 
the class of
solutions corresponding to this region presents similar properties to
the case of Fig.~\ref{plots-rs.75}.  The solutions for $R/q
=4\sqrt{3}/9\simeq 0.77$ are regular black holes for all $r_0/q$ ($0<
r_0/q \leq R/q$).

The boundary surface $r_0$ of the matter is timelike for all $r_0/q< R/q$,
and is lightlike for $r_0/q=R/q$. The nonextremal regular black holes with
$r_0/q< R/q$ (and $r_0/q\neq 2/3$) all have the same causal structure as 
shown in the conformal
diagram of Fig.~\ref{pen-timelk}, and, for values of $r_0/q$
sufficiently close to $R/q$,  the matter content of these regular black
holes satisfy a phantom-type equation of state.

The regular black holes with a lightlike boundary at $r_0/q=R/q$ are those
with a de Sitter core and a charged layer at $r_0$. These were studied in
detail in Sec.~\ref{sec-regdeSitterC}.

For $r_0/q=2/3$ the event and the Cauchy horizons coincide and we find
a regular extremal black hole although it has negative energy density
in some regions in the central core and we do not dwell further on it.

\subsubsection{$ 4\sqrt{3}/9 < R/q < \infty$: Intermediate and low
mass-energy densities}

(i) $ 4\sqrt{3}/9 < R/q < 1$: Intermediate
energy densities

\begin{figure}[h]
\centerline{\includegraphics[scale=.9]{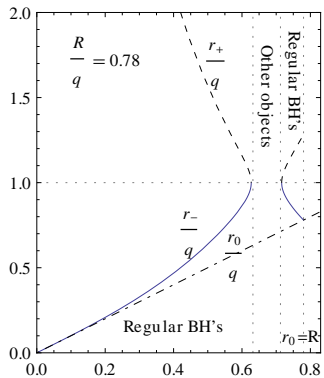}}
\caption{A plot of $r_+/q$ and $r_-/q$ as a function of $r_0/q$ 
for  $R/q = 0.78$. 
The first vertical dashed line is at $r_0/q\simeq0.63$, and the
second one is at $r_0/q\simeq 0.71$.}
\label{plots-rs.78}
\end{figure}

For more disperse, not so compact, charged mass distributions,
$R/q$ is in the range $ 4\sqrt{3}/9 < R/q < 1$.
A typical case in this range is shown in Fig.~\ref{plots-rs.78} for $R/q=
0.78$.

In Fig.~\ref{plots-rs.78} we plot the horizon radius $r_+$ and the Cauchy
horizon $r_-$ in terms of $r_0/q$ for $R/q= 0.78$.
There are regular black holes for some values of $r_0/q$.

The behavior of the mass function $m(r_0/q, R/q)$ is similar to the curve
for $R/q = 1$ in Fig.~\ref{plot-mass}.  Given a value of $R/q$ within 
this interval, the minimum of the corresponding
curve is at $\bar
r_0/q=\sqrt[4]{R^2/3q^2}$, which is in the interval $2/3\leq \bar r_0/q\leq
\sqrt[4]{3}$ and the values of the mass are bounded from below by $\bar
m/q\simeq 0.877$.

The regular black holes in the region $0<r_0/q < 0.63$ suffer from the
drawback of presenting negative energy density in some region
inside matter, i.e., $\rho_{\rm m}(r)$ assumes negative values for some $r$.
For $r_0/q$ in the interval $0.63 < r_0/q < 0.71$ the solutions correspond to
overcharged stars (regular and in some cases singular). Finally, in the
last region, $0.71< r_0/q < 1$, well-behaved regular black holes are
found with positive $\rho_{\rm m}(r)$ and phantom matter.

\vskip 0.2cm
(ii) {$R/q=1$: A special case with intermediate mass-energy density}

An interesting special  intermediate mass-energy density
case is the one that has $R/q=1.0$.

\begin{figure}
\centerline{\includegraphics[scale=.99]{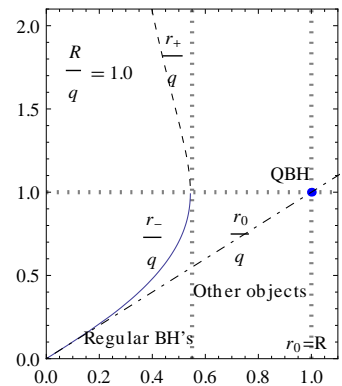}}
\caption{A plot of $r_+/q$ and $r_-/q$ as a function of $r_0/q$ 
for  $R/q=1.0$. The point indicated by QBH is the
quasiblack hole solution.}
\label{plots-rs1.0}
\end{figure}

In Fig.~\ref{plots-rs1.0}  we plot the horizon radius $r_+$ and the
Cauchy horizon $r_-$ in terms of $r_0/q$ for $R/q=1.0$.  The
regular black holes are found in these Guilfoyle solutions
in the region $0 < r_0/q\lesssim 0.55$, the largest matching
$r_0\sim 0.55$ occurs in the extremal black hole case, $r_-=r_+$. For 
solutions in this range the mass-energy density is negative in some 
regions inside
matter for solutions with small $r_0$ when compared to $R$. Hence, these 
are not good regular black holes.

Also, for 
$R/q=1$ in the limit $r_0/q \rightarrow R/q$ a
quasiblack hole is found. This is shown in Fig.~\ref{plots-rs1.0} by the
point indicated by QBH.
We comment further on this solution. 
As mentioned above, the maximum possible
value of $r_0$ is the de Sitter radius $R$, $r_0 =
R$.  In that limit, Eq.~\eqref{mass} implies $1-2m/r_0 +q^2/r_0^2=0$,
meaning that the boundary surface coincides with a horizon of the
exterior metric. More precisely, the matching is done at the inner
horizon of the exterior Reissner-Nordstr\"om spacetime, $r_0=r_-$,
with the event horizon $r_+$ being outside the matching surface.  The
only exception is the extremal case $m^2=q^2$, for which one has also
$r_-=r_+$, and then $r_0=r_+$. This case is represented by the point
QBH in Fig.~\ref{plots-rs1.0}, where the matching is at
$r_0/R=q^2/R^2=1$, and with $r_-=r_+$. In fact, this is not a regular
black hole. This is the quasiblack hole case \cite{lemoszanchin2010}.
The same type of solution was reported in
\cite{Lemos:2011vz}.

\vskip 0.2cm
(iii) {$1< R/q <\infty$: Low mass-energy densities}

\begin{figure}
\centerline{\includegraphics[scale=.85]{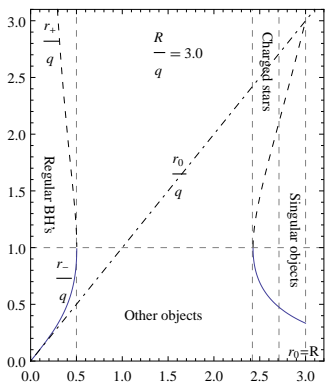}}
\caption{A plot of $r_+/q$ and $r_-/q$ as a function of $r_0/q$ 
for $R/q=3.0$, a typical small energy density  value.
}
\label{plots-rs3.0}
\end{figure}

For disperse, low mass-energy density distributions, $R/q$ is in the
range $ 1< R/q <\infty$. A typical case in this range is shown in
Fig.~\ref{plots-rs3.0} for $R/q=3.0$.

In Fig.~\ref{plots-rs3.0} we plot the horizon radius $r_+$ and the
Cauchy horizon $r_-$ in terms of $r_0/q$ for $R/q=3.0$. As in the case
of Fig.~\ref{plots-rs.78}, there are two branches for $r_+$ and $r_-$. One
branch ends at $r_0/q \simeq 0.51$ and the other starts at
$r_0/q\simeq 2.42$.  There are regular black holes for sufficiently
small $r_0/q$. The vertical dotted line at $r_0/q\simeq 0.51$ is drawn
to indicate that the regular black hole region ($r_0/q \leq r_-/q$) is
on the left of that point, i.e., there are regular black holes just
for $0< r_0/q \lesssim 0.51$. In the whole region $0.51\lesssim r_0/q
\leq R/q$ there are other kinds of objects, overcharged objects for $
0.51\lesssim r_0/q < 2.42 $, undercharged stars for $2.42\lesssim
r_0/q \lesssim 2.8$, and singular objects above this value, i.e., for
$2.8\lesssim r_0/q < 3.0$ only singular objects are found.

Note that the matching is at $r=r_0$ and then the curves for
$r_-(r_0,R)$ and $r_+(r_0,R)$ to the right of the vertical line at
$r_0/q\simeq 2.42$ in Fig.~\ref{plots-rs3.0} are not real horizons,
because they would be inside the matter.

When $r_0=R$  the boundary is at $r_0=R=r_+$, i.e., at 
the horizon. However, for $r_0$ close to $R$  the solutions 
are singular at one or several radii $r$. Thus, we 
do not analyze objects with $r_0$ close to $R$.

Again, the region where  regular black holes may be found
corresponds to relatively small values of the matter boundary radius
$r_0$.  This means that the pressure assumes very large negative
values in some regions, and, moreover, the energy density also becomes
negative in some regions inside matter. This drawback makes these
regular solutions less interesting than the ones occurring for large
energy density (with $R/q<1$).

\subsection{Comments}
\label{sec-furtherconstraints}

Two other interesting features should be mentioned.

\noindent
(i) The limit of zero $r_0$, for fixed $R/q$, gives singular charged
black holes.

\noindent
(ii) The limit of zero charge, with nonzero $r_0$, of these solutions
is not regular black holes, but corresponds to uncharged
stars for a given range of parameters. 
The exception is the limit of zero charge and zero
$r_0$ that gives a Schwarzschild black hole.

\subsection{On the numerical techniques employed}

The starting point for the analysis is a system of relations defining
the metric potentials $A(r)$ and $B(r)$ that define the spacetime
geometry, and the other quantities that comprise the energy-momentum
tensor; namely, the fluid quantities $\rho_m(r)$, $p(r)$, and
$U^\mu$, and the electromagnetic energy density $Q^2(r)/r^4$.  These
quantities are given in terms of three free parameters $r_0$, $q$, and
$R$ representing the radius, the electric charge, and a characteristic
length related to the total energy density of the fluid. After
normalizing with respect to $q$, we are left with two free parameters
from what the other important parameters such as the mass $m$,
horizons radii $r_+$ and $r_-$, 
and Guilfoyle parameter $a$ are obtained,
i.e., each nonfree parameter is treated as a function of the two free
parameters, $m=m(r_0/q,R/q)$, etc., while the potentials and fluid
quantities depend also on the radial coordinate $r$, $A(r)=
A(r_0/q,R/q,r/q)$, and so on.

Since we have analytical expressions for all these quantities, the related 
numerical techniques employed to investigate the behavior of each functions
are mainly visualization of graphics and approximation techniques, 
numerical or analytical.

The region where black holes are found is given by the conditions
$m^2> q^2$ and $r_0\leq r_-$.  Equations~\eqref{mass} and \eqref{rpm}
help to locate such a region. Then, after choosing a specific pair of
parameters $(r_0/q,R/q)$ satisfying these constraints, all the
relevant functions are analyzed for the required physical conditions,
in particular, the regularity of the energy density, pressure and
electromagnetic energy density for all values of the radial coordinate
$r$. The curvature scalars are also checked. This is done by using
standard methods of functional analysis.  We repeated the process by
varying the two free parameters $r_0/q$ and $R/q$ along the appropriate
region of the parameter space.

In summary, all the numerical and graphical analysis may be done using built 
in functions of the chosen algebraic software, e.g., using the internal 
functions of Wolfram Mathematica.

\section{Conclusions}
\label{sec-conclusion}

Charged regular black holes have been found here for a range of
parameters of the static spherically symmetric solutions displayed by
Guilfoyle.
For a range of parameters, the solutions are regular
electrically charged black hole solutions. They are built from charged
phantom matter satisfying the
condition $\rho_{\rm m}+p\leq 0$ up to $r_0$.  The metric for $r<r_0$ is
regular and the isotropic
pressure goes to $0$ at $r_0$ while the energy density and the charge
density may present discontinuities at $r_0$.
The exterior is the Reissner-Nordstr\"om solution.
The matter boundary is timelike
and is always
at a radius smaller than the
inner Reissner-Nordstr\"om Cauchy horizon.

In the limiting case where $ \rho_{\rm m}+p= 0$ the interior can be
interpreted as a false vacuum de Sitter state. In this case, the
isotropic pressure and the energy density are constant throughout the
interior region.
The boundary is lightlike, located at the
inner Reissner-Nordstr\"om Cauchy horizon $r_-$.
All the electric charge 
of the solution is located at this boundary.

\section*{Acknowledgments}

JPSL thanks Funda\c c\~ao para a Ci\^encia e 
Tecnologia (FCT), Portugal, for financial support through Grant
No.~UID/FIS/00099/2013.  VTZ thanks Funda\c c\~ao Calouste
Gulbenkian for financial support.  VTZ also thanks Funda\c c\~ao de
Amparo \`a Pesquisa do Estado de S\~ao Paulo (FAPESP) and Conselho
Nacional de Desenvolvimento Cient\'\i fico e Tecnol\'ogico of Brazil
(CNPq) for financial help. We thank Coordena\c{c}\~ao de
Aperfei\c{c}oamento do Pessoal de N\'\i vel Superior (CAPES), Brazil, Grant
No.~88881.064999/2014-01.

\end{document}